\documentclass[12pt,preprint]{aastex}
\usepackage{graphicx}
\usepackage{aastexug}

\shorttitle{A TOY MODEL FOR GAMMA-RAY BURSTS IN TYPE IB/C
SUPERNOVAE} \shortauthors{D. X.Wanget al.}

\begin{document}
\title{A TOY MODEL FOR GAMMA-RAY BURSTS IN TYPE IB/C SUPERNOVAE}

\author{WEI-HUA LEI, DING-XIONG WANG\altaffilmark{1}, AND REN-YI
MA} \affil{Department of Physics, Huazhong University of Science
and Technology, Wuhan, 430074, People's Republic of China}
\email{dxwang@hust.edu.cn}

\altaffiltext{1}{Send offprint requests to: Ding-Xiong Wang
(dxwang@hust.edu.cn)}

\begin{abstract}
A toy model for gamma-ray burst-supernovae (GRB-SN) is discussed
by considering the coexistence of baryon poor outflows from black
holes (BHs) and a powerful spin-connection to surrounding disk,
giving rise to consistent calorimetry as described by van Putten
(2001a) in a variant of the Blandford-Znajek process (BZ, 1977).
In this model the half-opening angle of the magnetic flux tube on
the horizon is determined by the mapping relation between the
angular coordinate on the BH horizon and the radial coordinate on
the surrounding accretion disk. The GRB is powered by the baryon
poor outflows in the BZ process, and the associated SN is powered
by very small fraction of the spin energy transferred from the BH
to the disk in the magnetic coupling (MC) process. The timescale
of the GRB is fitted by the duration of the open magnetic flux on
the horizon. It turns out that the data of several GRB-SNe are
well fitted with our model.
\end{abstract}


\section{INTRODUCTION}

As is well known, the Blandford-Znajek (BZ) process is an
effective mechanism for powering jets from quasars and active
galactic nuclei (Blandford {\&} Znajek 1977; Rees 1984). In the BZ
process energy and angular momentum are extracted from a rotating
balck hole (BH), and transferred to the remote astrophysics load
by the open magnetic field lines. Recently, much attention has
been paid on the issue of the long duration gamma-ray bursts
(GRBs) powered by the BZ process (Paczynski 1993, 1998; Meszaros
{\&} Rees 1997). A detailed model of GRBs invoking the BZ process
is given by Lee, Wijers, {\&} Brown 2000, hereafter L00), and they
concluded that a fast-spinning BH with strong magnetic field $\sim
10^{15}G$ can provide energy $\sim 10^{53}ergs$ to power a GRB
within 1000 seconds.

Recently the observations and theoretical considerations have
linked long duration GRB with ultra-bright type Ib/c supernovae
(Galama et al. 1998; Bloom et al. 1999; Galama et al. 2000). The
first candidate was provided by SN1998bw and GRB980425, and the
recent HETE-II burst GRB030329 has greatly enhanced the confidence
in this association (Stanek et al. 2003; Hjorth et al. 2003).

Not long ago Brown et al. (2000, hereafter B00) worked out a
specific scenario for GRB-SN connection. They argued that the GRB
is powered by the BZ process, and the SN is powered by the energy
dissipated into the disk through closed magnetic field lines
coupling the disk with the BH. The latter energy mechanism is
referred to as the magnetic coupling (MC) process, which is
regarded as one of the variants of the BZ process (Blandford 1999;
Putten 1999; Li 2000, 2002; Wang, Xiao {\&} Lei 2002, hereafter
W02). It is shown in B00 that about $10^{53}ergs$ are available to
power both GRBs and SNe. However, they failed to distinguish the
fractions of the energy for these two objects.

More recently, van Putten et al. (van Putten 2001a; van Putten \&
Levinson 2003; van Putten et al. 2004) worked out a poloidal
topology for the open and closed magnetic field lines, where the
separatrix on the horizon is defined by a finite half-opening
angle. The duration of GRB is set by the lifetime of rapid spin of
the BH. It is found that GRB and SN are powered by a small
fraction of the BH spin energy. This result is consistent with the
observations, i.e., the duration of GRB with tens of seconds, the
true GRB energies distributed around $5\times 10^{50}ergs$ (Frail
et al. 2002), and the aspherical SNe kinetic energies with
$2\times 10^{51}ergs$ ( Hoflich et al. 1999).

In this paper we propose a toy model for GRB-SN by considering
coexistence of the BZ and MC processes (CEBZMC), where the
configuration of the magnetic field is based the works of Wang et
al. and van Putten et al. (W02; Wang et al. 2003, hereafter W03;
van Putten 2001a; van Putten et al. 2004). The transferred energy
and the duration for GRBs are calculated in the evolving process
of the half-opening angle of the magnetic flux on the BH horizon,
while the transferred energy for SNe is calculated in the MC
process. Thus the association of GRBs and SNe is explained
reasonably and consistently.

This paper is organized as follows. In \S \ 2 the configuration of
the magnetic field is described with the half-opening angle
$\theta _{BZ} $ of the magnetic flux tube on the horizon, and we
find that $\theta _{BZ} $ can evolve to zero in some range of the
power-law index $n$ of the magnetic field on the disk. In \S \ 3
the toy model for GRB-SN based on BH evolution is described in the
parameter space consisting of the BH spin $a_ * $ and the
power-law index $n$. It is shown that the half-opening angle
$\theta _{BZ} $ plays an important role of determining (i) the
powers for GRBs and SNe, (ii) the duration of GRB. In \S \ 4 we
calculate the duration of GRB and the energies of GRB-SN, and
compare with the other models of GRBs invoking the BZ process. It
is found the duration of GRB obtained in our model is shorter than
that of the other models. It turns out that this model is in
excellent agreement with the observations, owing to considering
the MC effects, while too much energy than the needed energy of
GRB is produced in the other models without the MC effects.
Finally, in \S \ 5, we summarize our main results and discuss some
problems concerning our model. Throughout this paper, the
geometric units $G = c = 1$ are used.

\section{CONFIGURATION OF THE MAGNETIC FIELD }

Recently, we proposed a model of CEBZMC, in which the remote
astrophysical load in the BZ process and the disk load in the MC
process are connected with a rotating BH by open and closed
magnetic field lines, respectively (W02; W03). The poloidal
configuration of the magnetic field is shown in Figure 1, which is
adapted from van Putten (2001a).

In Figure 1 the angle $\theta _{BZ} $ is the half-opening angle of
the open magnetic flux tube, indicating the angular boundary
between open and closed field lines on the horizon. This angle has
been discussed by van Putten and his collaborators, who related
the half-opening angle to the curvature in poloidal topology of
the inner torus magnetosphere (van Putten {\&} Levinson 2003; van
Putten et al. 2004). In W03 the angle $\theta _{BZ} $ is
determined based on the following assumptions.

(\ref{eq1}) The theory of a stationary, axisymmetric magnetosphere
formulated in the work of MacDonald {\&} Thorne (1982) is
applicable not only to the BZ process but also to the MC process.
The magnetosphere is assumed to be force-free outside the BH and
disk.

(\ref{eq2}) The disk is both stable and perfectly conducting, and the closed
magnetic field lines are frozen in the disk. The disk is thin and Keplerian,
and it lies in the equatorial plane of the BH, with inner boundary being at
the marginally stable orbit.

(\ref{eq3}) The poloidal magnetic field is assumed to be constant
on the horizon and to vary as a power law on the disk as follows
(Blandford 1976),

\begin{equation}
\label{eq1}
B_D \propto \xi ^{ - n},
\end{equation}

\noindent
where $B_D $ is the magnetic field on the disk, the parameter $n$ is the
power-law index, and $\xi \equiv r / r_{ms} $ is the radial coordinate on
the disk, which is defined in terms of the radius $r_{ms} \equiv M\chi
_{ms}^2 $ of the marginally stable orbit (Novikov {\&} Thorne 1973).

(\ref{eq4}) The magnetic flux connecting the BH with its surrounding disk takes
precedence over that connecting the BH with the remote load.

Assumption (\ref{eq4}) is proposed based on two reasons: (i) the
magnetic field on the horizon is brought and held by the
surrounding magnetized disk, and (ii) the disk is much nearer to
the BH than the remote load.

A mapping relation between the parameter $\xi $ and the angular
$\theta $ is derived in W03 based on the conservation of magnetic
flux of the closed field lines.

\begin{equation}
\label{eq2}
\cos \theta = \int_1^\xi {\mbox{G}\left( {a_ * ;\xi ,n} \right)d\xi } ,
\end{equation}

\noindent
where $a_\ast \equiv J / M^2$ is the BH spin defined by the BH mass $M$ and
angular momentum $J$, and the function $G(a_\ast ;\xi ,n)$ is given by

\begin{equation}
\label{eq3}
\mbox{G}\left( {a_ * ;\xi ,n} \right) = \frac{\xi ^{1 - n}\chi _{ms}^2 \sqrt
{1 + a_ * ^2 \chi _{ms}^{ - 4} \xi ^{ - 2} + 2a_ * ^2 \chi _{ms}^{ - 6} \xi
^{ - 3}} }{2\sqrt {\left( {1 + a_ * ^2 \chi _{ms}^{ - 4} + 2a_ * ^2 \chi
_{ms}^{ - 6} } \right)\left( {1 - 2\chi _{ms}^{ - 2} \xi ^{ - 1} + a_ * ^2
\chi _{ms}^{ - 4} \xi ^{ - 2}} \right)} }.
\end{equation}

Based on the assumption (\ref{eq4}) the angle $\theta _{BZ} $ can be determined by
taking $\xi = \infty $ for the highest closed field lines as follows,

\begin{equation}
\label{eq4}
\cos \theta _{BZ} = \int_1^\infty {\mbox{G}\left( {a_ * ;\xi ,n} \right)d\xi
} .
\end{equation}

The curves of $\theta _{BZ} $ versus $a_\ast $ with different values of
power-law index $n$ are shown in Figure 2.

Inspecting Figure 2, we have the following results.

(\ref{eq1}) The half-opening angle $\theta _{BZ} $ always increases monotonically
with the increasing $n$.

(\ref{eq2}) For some values of $n$, such as $n$= 3.5, 4, 4.5, 5 and 5.493, the angle
$\theta _{BZ} $ can evolve to zero with the decreasing $a_\ast $.

The second point implies that the open magnetic flux tube will be
shut off when the BH spin decreases to the critical value $a_ *
^{GRB} $ corresponding to $\theta _{BZ} = 0$. The lifetime of the
half-opening angle $\theta _{BZ} $ is defined as the evolution
time of BH from the initial spin $a_\ast (0)$ to $a_ * ^{GRB} $.
Obviously it is less than the lifetime of BH spin, which is the
evolution time of BH from $a_\ast (0)$ to zero. By using equation
(\ref{eq4}) we find that $a_\ast ^{GRB} $ exists only for a
specific value range of the power-law index $n$, i.e., $3.003 \le
n \le 5.493$. The half-opening angle $\theta _{BZ} $ remains zero
for $n < 3.003$ in the whole evolving process, and it will never
evolve to zero for $n > 5.493$.

\section{A TOY MODEL FOR GRB-SN BASED ON BH EVOLUTION IN CEBZMC}

Considering that the angular momentum is transferred from a
rapidly rotating BH to the disk, on which a positive torque
exerts, we think the accretion onto the BH might be probably
halted. This state is essentially the suspended accretion state
proposed by van Putten {\&} Ostriker (2001), and the evolution of
the BH is governed by the BZ and the MC processes.

The expressions for the BZ and MC powers and torques are derived in W02 by
using an improved equivalent circuit for the BH magnetosphere based on the
work of MacDonald and Thorne (1982). Considering the angular boundary
$\theta _{BZ} $, we express the BZ and MC powers and torques as follows.

\begin{equation}
\label{eq5}
\tilde {P}_{BZ} \equiv {P_{BZ} } \mathord{\left/ {\vphantom {{P_{BZ} } {P_0
}}} \right. \kern-\nulldelimiterspace} {P_0 } = 2a_ * ^2 \int_0^{\theta
_{BZ} } {\frac{k\left( {1 - k} \right)\sin ^3\theta d\theta }{2 - \left( {1
- q} \right)\sin ^2\theta }} ,
\end{equation}

\begin{equation}
\label{eq6}
{\tilde {T}_{BZ} \equiv T_{BZ} } \mathord{\left/ {\vphantom {{\tilde
{T}_{BZ} \equiv T_{BZ} } {T_0 }}} \right. \kern-\nulldelimiterspace} {T_0 }
= 4a_ * \left( {1 + q} \right)\int_0^{\theta _{BZ} } {\frac{\left( {1 - k}
\right)\sin ^3\theta d\theta }{2 - \left( {1 - q} \right)\sin ^2\theta }} ,
\end{equation}

\begin{equation}
\label{eq7}
{\tilde {P}_{MC} \equiv P_{MC} } \mathord{\left/ {\vphantom {{\tilde
{P}_{MC} \equiv P_{MC} } {P_0 }}} \right. \kern-\nulldelimiterspace} {P_0 }
= 2a_ * ^2 \int_{\theta _{BZ} }^{\pi / 2} {\frac{\beta \left( {1 - \beta }
\right)\sin ^3\theta d\theta }{2 - \left( {1 - q} \right)\sin ^2\theta }} ,
\end{equation}

\begin{equation}
\label{eq8}
{\tilde {T}_{MC} \equiv T_{MC} } \mathord{\left/ {\vphantom {{\tilde
{T}_{MC} \equiv T_{MC} } {T_0 }}} \right. \kern-\nulldelimiterspace} {T_0 }
= 4a_ * \left( {1 + q} \right)\int_{\theta _{BZ} }^{\pi / 2} {\frac{\left(
{1 - \beta } \right)\sin ^3\theta d\theta }{2 - \left( {1 - q} \right)\sin
^2\theta }} ,
\end{equation}

\noindent
where we have $q \equiv \sqrt {1 - a_\ast ^2 } $ and

\begin{equation}
\label{eq9}
\left\{ {\begin{array}{l}
 P_0 \equiv \left\langle {B_H^2 } \right\rangle M^2 \approx 6.59\times
10^{50}\times \left( {\frac{B_H }{10^{15}G}} \right)^2\left( {\frac{M}{M_
\odot }} \right)^2erg \cdot s^{ - 1} \\
 T_0 \equiv \left\langle {B_H^2 } \right\rangle M^3 \approx 3.26\times
10^{45}\left( {\frac{B_H }{10^{15}G}} \right)^2\left( {\frac{M}{M_ \odot }}
\right)^3g \cdot cm^2 \cdot s^{ - 2} \\
 \end{array}} \right.
\end{equation}

In equation (\ref{eq9}) $B_H $ is the magnetic field on the BH horizon. The
parameters $k$ and $\beta $ are the ratios of the angular velocities of the
open and closed magnetic field lines to that of the BH, respectively.
Usually, $k = 0.5$ is taken for the optimal BZ power. Since the closed field
lines are assumed to be frozen in the disk, the ratio $\beta $ is related to
$a_\ast $ and $\xi $ by

\begin{equation}
\label{eq10}
\beta \equiv {\Omega _D } \mathord{\left/ {\vphantom {{\Omega _D } {\Omega
_H }}} \right. \kern-\nulldelimiterspace} {\Omega _H } = \frac{2\left( {1 +
q} \right)}{a_ * \left[ {\left( {\sqrt \xi \chi _{ms} } \right)^3 + a_ * }
\right]},
\end{equation}

\noindent
where $\Omega _H $ and $\Omega _D $ are the angular velocities of the BH and
the disk, respectively.

Since $\theta _{BZ} $ in equations (\ref{eq5}) and (\ref{eq7}) is determined by $a_\ast $
and $n$, we have the curves of the powers $\tilde {P}_{BZ} $ and $\tilde
{P}_{MC} $ versus $a_\ast $ for the given values of $n$ as shown in Figure
3.

Inspecting Figure 3, we have the following results.

(\ref{eq1}) The MC power $P_{MC} $ is always greater than the BZ
power $P_{BZ} $ for a wide value range of the power-law index $n$,
provided that the BH spin $a_ * $ is not very small.

(\ref{eq2}) For the given values of $a_\ast $ the BZ power $P_{BZ} $ increases with
the increasing $n$, while the MC power $P_{MC} $ decreases with it.

(\ref{eq3}) The power $P_{MC} = 0$ holds at some small values of $a_\ast $ for the
given values of the index $n$.

From equation (\ref{eq5}) we find $P_{BZ} = 0$ when $\theta _{BZ}
= 0$. Therefore the effective time of the BZ process can be set by
the lifetime of $\theta _{BZ} $, i.e., the evolution time of BH
from $a_\ast (0)$ to $a_\ast ^{GRB} $ for $3.003 \le n \le 5.493$.
As argued in our previous work (Wang, Lei {\&} Ma 2003), we have
$P_{MC} = 0$ when the energy transported from the BH to the disk
is equal to that from the disk to the BH. Henceforth the
corresponding critical BH spin is denoted by $a_\ast ^{SN} $. We
have $P_{MC} > 0$ for $a_\ast > a_\ast ^{SN} $, and $P_{MC} < 0$
for $a_\ast < a_\ast ^{SN} $, indicating that the energy
transferred from the BH to the disk dominates that transferred in
the inverse direction.

Based on the conservation laws of energy and angular momentum we have the
following evolution equations of the rotating BH,

\begin{equation}
\label{eq11}
{dM} \mathord{\left/ {\vphantom {{dM} {dt}}} \right.
\kern-\nulldelimiterspace} {dt} = - (P_{BZ} + P_{MC} ),
\end{equation}

\begin{equation}
\label{eq12}
{dJ} \mathord{\left/ {\vphantom {{dJ} {dt}}} \right.
\kern-\nulldelimiterspace} {dt} = - (T_{BZ} + T_{MC} ).
\end{equation}

Incorporating equations (\ref{eq11}) and (\ref{eq12}), we have the evolution equation for
the BH spin expressed by

\begin{equation}
\label{eq13}
{da_ * } \mathord{\left/ {\vphantom {{da_ * } {dt}}} \right.
\kern-\nulldelimiterspace} {dt} = - M^{ - 2}(T_{BZ} + T_{MC} ) + 2M^{ - 1}a_
* (P_{BZ} + P_{MC} ) = B_H^2 MA(a_\ast ,n)
\end{equation}

\noindent
where the function $A(a_\ast ,n)$ is

\begin{equation}
\label{eq14}
A(a_\ast ,n) = 2a_\ast \tilde {P}_{mag} - \tilde {T}_{mag} ,
\end{equation}

\noindent
where

\begin{equation}
\label{eq15}
\tilde {P}_{mag} = \tilde {P}_{BZ} + \tilde {P}_{MC} ,\mbox{ }\tilde
{T}_{mag} = \tilde {T}_{BZ} + \tilde {T}_{MC} .
\end{equation}

The sign of $A(a_\ast ,n)$ determines whether $a_\ast $ decreases or
increases, and we have the curves of $A(a_\ast ,n)$ versus $a_\ast $ for
different values of $n$ as shown in Figure 4.

From Figure 4, we find the following evolution characteristics of
the BH spin.

(\ref{eq1}) The function $A(a_\ast ,n) > 0$ for $a_ * < a_\ast
^{eq} $, and $A(a_\ast ,n) < 0$ for $a_ * > a_\ast ^{eq} $, and
the BH spin always evolve to the equilibrium spin $a_\ast ^{eq} $.

(\ref{eq2}) Both the equilibrium spin $a_\ast ^{eq} $ and the
decreasing rate of the BH spin are sensitive to the power-law
index $n.$ The greater the index $ n$ is, the greater is the
equilibrium spin $a_\ast ^{eq} $, and the more slowly the BH spin
decreases to $a_\ast ^{eq} $.

(\ref{eq3}) The features of our model for GRB-SN can be described
by three functions: $A(a_\ast ,n)$, $\tilde {P}_{BZ} (a_\ast ,n)$
and $\tilde {P}_{MC} (a_\ast ,n)$.

Setting $A(a_\ast ,n) = 0$, $\tilde {P}_{BZ} (a_\ast ,n) = 0$ and
$\tilde {P}_{MC} (a_\ast ,n) = 0$, we have the three
characteristic curves in the parameter space as shown in Figure 5.

The parameter space is divided into seven regions, \textbf{IA},
\textbf{IB}, \textbf{II}, \textbf{IIIA},\textbf{ IIIB},
\textbf{IVA} and \textbf{IVB}, where the points \textbf{\emph{M}}
and \textbf{\emph{N}} are the intersections of the curve $\tilde
{P}_{BZ} (a_\ast ,n) = 0$ with the curves $\tilde {P}_{MC} (a_\ast
,n) = 0$ and $A(a_\ast ,n) = 0$, respectively. \textbf{IA} and
\textbf{IB} is divided by the dashed line starting from the point
\textbf{\emph{M}}. By using the equations (\ref{eq5}), (\ref{eq7})
and (\ref{eq14}) we obtain the coordinates of points
\textbf{\emph{M}} and \textbf{\emph{N}}, i.e.,
\textbf{\emph{M}}(\textbf{0.253, 5.067}) and
\textbf{\emph{N}}(\textbf{0.222, 5.122}), respectively. In Figure
5 each black dot with an arrowhead in these sub-regions is
referred to as a representative point (\textbf{RP}), which
represents one evolution state of the BH as shown in Table 1.

It is found from Figure 5 and Table 1 that GRB-SN only occurs in
region \textbf{I } and \textbf{I } corresponding to CEBZMC. Based
on the observations that GRBs need less energy than SNe (B00), we
think that the evolving path of the\textbf{ RP} in region
\textbf{IA} is reasonable, which corresponds to $3.003 < n <
5.067$, and the RP attains $\tilde {P}_{BZ} (a_\ast ,n) = 0$ with
$\tilde {P}_{MC} (a_\ast ,n) > 0$ in braking the spinning BH.

Based on the above discussion we describe the toy model for GRB-SN
in the parameter space as follows.

(\ref{eq1}) The fast-spinning BH can power GRBs and SNe via the BZ
and MC processes, respectively, which is represented by the
\textbf{RP} in region \textbf{IA}. The GRB will be shut off, when
the\textbf{ RP }reaches eventually the curve $\tilde {P}_{BZ}
(a_\ast ,n) = 0$.

(\ref{eq2}) Energy is transferred from the spinning BH to the disk
via the MC process in the evolving path from region \textbf{IA} to
region \textbf{II}. The SN will be shut off with no further energy
transferred into the disk, if the \textbf{RP} arrives at the curve
$\tilde {P}_{MC} (a_\ast ,n) = 0$.

(\ref{eq3}) Since enormous energy is deposited in the disk due to
the MC process, the disk might be destroyed eventually, leaving
the BH alone without the disk. This outcome probably arises from
explosion of SNe (B00).

\section{ENERGIES AND TIME SCALES FOR GRB-SN}

In the evolving path of the \textbf{RP} going through regions
\textbf{IA}, \textbf{II} and \textbf{IIIA} we have $a_* ^{GRB} \ge
a_* ^{SN} > a_* ^{eq} $. Based on the above discussion on
correlation of the BH evolution with the association of GRB-SN ,
the energy $E_{BZ} $ and $E_{SN} $ extracted in the BZ and MC
processes are given respectively by

\begin{equation}
\label{eq16}
E_{BZ} = \int_{a_\ast (0)}^{a_ * ^{GRB} }
{\frac{P_{BZ} }{\left( {{da_ * } \mathord{\left/ {\vphantom {{da_
* } {dt}}} \right. \kern-\nulldelimiterspace} {dt}} \right)}} da_
* = 1.79\times 10^{54}ergs\times \left( {\frac{M(0)}{M_ \odot }}
\right)\int_{a_\ast (0)}^{a_ * ^{GRB} } {\frac{\tilde {M}\tilde
{P}_{BZ} }{ - \tilde {T}_{mag} + 2a_ * \tilde {P}_{mag} }} da_ * ,
\end{equation}

\begin{equation}
\label{eq17}
E_{MC} = \int_{a_\ast (0)}^{a_ * ^{SN} }
{\frac{P_{MC} }{\left( {{da_ * } \mathord{\left/ {\vphantom {{da_
* } {dt}}} \right. \kern-\nulldelimiterspace} {dt}} \right)}} da_
* = 1.79\times 10^{54}ergs\times \left( {\frac{M(0)}{M_ \odot }}
\right)\int_{a_\ast (0)}^{a_ * ^{SN} } {\frac{\tilde {M}\tilde
{P}_{MC} }{ - \tilde {T}_{mag} + 2a_ * \tilde {P}_{mag} }} da_ * .
\end{equation}

The true energy for GRBs, $E_\gamma $, and the energy for SNe,
$E_{SN} $, are related respectively to $E_{BZ} $ and $E_{MC} $ by

\begin{equation}
\label{eq18} E_\gamma = \varepsilon _\gamma E_{BZ} ,
\end{equation}

\begin{equation}
\label{eq19} E_{SN} = \varepsilon _{SN} E_{MC} ,
\end{equation}

\noindent where $\varepsilon _\gamma $ and $\varepsilon _{SN} $
denote the efficiencies of converting $E_{BZ} $ and $E_{MC}$ into
$E_\gamma $ and $E_{SN} $, respectively. Following van Putten et
al. (2004), we take $\varepsilon _\gamma = 0.15$ in calculations.
The duration of GRB, $t_{GRB} $, is defined as the lifetime of the
half-opening angle $\theta _{BZ} $, which is exactly equal to the
time for the BH spin evolving from $a_\ast (0)$ to $a_\ast ^{GRB}
$, i.e.,

\begin{equation}
\label{eq20}
t_{GRB} = 2.7\times 10^3s\times \left(
{\frac{10^{15}G}{B_H }} \right)^2\left( {\frac{M_ \odot }{M(0)}}
\right)\int_{a_\ast (0)}^{a_\ast ^{GRB} } {\frac{\tilde {M}^{ -
1}}{ - \tilde {T}_{mag} + 2a_ * \tilde {P}_{mag} }da_ * } ,
\end{equation}

\noindent
where $\tilde {M}$ is the ratio of the BH mass to its
initial value $M(0)$, and it is calculated by

\begin{equation}
\label{eq21}
\tilde {M} \equiv \frac{M}{M(0)} = \exp \int_{a_\ast
(0)}^{a_ * } {\frac{da_
* }{{\tilde {T}_{mag} } \mathord{\left/ {\vphantom {{\tilde {T}_{mag} }
{\tilde {P}_{mag} }}} \right. \kern-\nulldelimiterspace} {\tilde
{P}_{mag} } - 2a_ * }} .
\end{equation}

\noindent Equation (\ref{eq21}) can be derived by using equations
(\ref{eq11}) and (\ref{eq13}). Obviously, $t_{GRB} $ is shorter
than the lifetime of the BH spin evolving from $a_\ast (0)$ to
$a_\ast = 0$.

Considering that spin-energy of the BHs produced in core-collapse
is around~50{\%} of maximum or less in centered nucleation (van
Putten 2004), we take the initial BH spin as $a_\ast (0) = 0.9$ in
equations (\ref{eq16}), (\ref{eq17}) and (\ref{eq21}). The value
range $3.534 < n < 5.067$ is taken in the evolving path of the
\textbf{RP} in region \textbf{IA }corresponding to GRB-SN. In
addition, $M(0) = 7M_ \odot $ and $B_H = 10^{15}G$ are assumed,
and the cutoff of $T_{GRB} $ is taken as $T_{90} $ in
calculations, which is the duration for 90{\%} of the total BZ
energy to be extracted (Lee {\&} Kim 2002).

As is well know, the rotational energy $E_{rot} $ of a Kerr BH is
expressed by (Thorne, Price {\&} Macdonald 1986)

\begin{equation}
\label{eq22}
E_{rot} = f(a_\ast )M(0),
\end{equation}

\noindent
where

\begin{equation}
\label{eq23}
f(a_\ast ) = 1 - \sqrt {\textstyle{1 \over 2}(1 +
\sqrt {1 - a_\ast ^2 } )}.
\end{equation}

\noindent
Substituting $a_\ast (0) = 0.9$ and $M(0) = 7M_ \odot $
into equation (\ref{eq22}), we have the total rotational energy of
the BH,

\begin{equation}
\label{eq24}
E_{rot} \approx 1.9\times 10^{54}ergs.
\end{equation}

To compare rotational energy of the BH extracted in the BZ and MC
processes, we calculate the ratios of $E_{BZ} $ and $E_{MC} $ to
$E_{rot} $, i.e.,

\begin{equation}
\label{eq25}
\eta _{BZ} \equiv E_{BZ} / E_{rot} , \quad \eta _{MC}
\equiv E_{MC} / E_{rot} .
\end{equation}

\noindent
The curves of $\eta _{BZ} $ and $\eta _{MC} $ versus $n$
are shown in Figure 6.

From Figure 6 we find that $E_{MC} $ is significantly greater than
$E_{BZ} $, and the greater the index$ n$, the more energy
extracted by BZ process. These results can be also obtained by
inspecting Figures 3 and 5. The ratio $\eta _{MC} $ varies from
25.8{\%} to 33.8{\%} with the ratio $\eta _{BZ} $ varying from
zero to 3.7{\%} for $3.534 < n < 5.067$. Combining equations
(\ref{eq24}) and (\ref{eq25}) with the above ratios, we obtain
$E_{MC} $ varying from $4.929\times 10^{53}ergs$ to $6.459\times
10^{53}ergs$ with $E_{BZ} = 7.003\times 10^{52}ergs$. And the
energy for GRB  is inferred by using $\varepsilon _\gamma = 0.15$.

Calculations of SN1998bw with aspherical geometry show that the
needed kinetic energy is about $2\times 10^{51}ergs$ (Hoflich et
al. 1999), and this results in the ratio $\varepsilon _{SN} $
varying from 0.003 to 0.004 for $E_{MC} $ varying from
$6.459\times 10^{53}ergs$ to $4.929\times 10^{53}ergs$ in equation
(\ref{eq19}). In this paper we take the average value,
$\varepsilon _{SN} \approx 0.0035$.

It has been shown that most fraction of the rotational energy of a
Kerr BH is emitted in unseen channels, such as in gravitational
radiation and MeV-neutrino emissions (van Putten 2001a, 2001b; van
Putten and Levinson 2002). The above efficiency factors obtained
in our model, $\varepsilon _\gamma = 0.15$ and $\varepsilon _{SN}
\approx 0.0035$, imply that only very small fraction of $E_{rot} $
is converted into the energy for GRBs and SNe. These results are
consistent with those obtained by van Putten and his
collaborators.

By using equation (\ref{eq20}) the duration of GRBs, $T_{90} $,
versus the parameter $n$ is shown in Figure 7.

We find that the estimated duration of GRB is tens of seconds,
which is consistent with the observations (Kouveliotous et al.
1993). For $n = 4$ and $B_H = 1\times 10^{15}G$ in the evolving
path of the \textbf{RP} from $a_\ast (0) = 0.9$ to $a_\ast ^{SN} =
0.235$, we obtain the duration of GRB $T_{90} = 21s$, and the
energies for GRB and SN are $E_\gamma = 5\times
10^{50}ergs(\textstyle{{\varepsilon _\gamma } \over {0.15}})$ and
$E_{SN} = 2\times 10^{51}ergs(\textstyle{{\varepsilon _{SN} }
\over {0.0035}})$, respectively. These results are in excellent
agreement with the observed durations of tens of seconds
(Kouveliotous et al. 1993), energies $E_\gamma = 5\times
10^{50}ergs$ in gamma rays (Frail et al. 2002), and inferred
kinetic energy $E_{SN} = 2\times 10^{51}ergs$ in SN1998bw with
aspherical geometry (Hoflich et al. 1999).

For several GRB-SNe, the observed energy $E_\gamma $ and duration
$T_{90} $ can be fitted by adjusting the parameters $n$ and $B_H
$, and the energy $E_{SN} $ can be predicted as shown in Table 2.

In the previous model for GRBs powered by the BZ process, the
duration of GRB is estimated either in the case that the BH is
spun down to zero or in the case that the whole disk is plunged
into the BH (L00; Lee {\&} Kim 2000, 2002; Wang, Lei {\&} Xiao
2002). However, the MC effects have not been taken into account in
these models. In this model the duration of GRB is estimated by
the lifetime of the half-opening angle on the BH horizon based on
CEBZMC. It turns out that both the BZ and MC processes play very
important roles for GRN-SN in this model.

The quantities, $E_\gamma $ and $T_{90} $, calculated in the three
different models for GRBs are listed in Table 3, where the
abbreviations\textbf{ CEBZMC, BZO }and \textbf{BZACC }represent
this model, the model invoking the BZ process only, the model
invoking the BZ process with disk accretion, respectively.

It is found in Table 3 that the energy and duration time obtained
in this model are in excellent agreement with the observed
durations of tens of seconds and energies $E_\gamma \approx
5\times 10^{50}ergs$. However the energies obtained in the models
\textbf{BZO} and \textbf{BZACC} seem too much to fit the
observations.

\section{DISCUSSION}

In this paper we discuss a toy model for GRB-SN, where the energy
for GRBs and SNe are powered by the BZ and MC processes,
respectively, and the duration for GRB is estimated by the
lifetime of the half-opening angle $\theta _{BZ} $. The energy
extracted in the BZ process for GRBs is much less than that
extracted in the MC process for SNe. This result is consistent
with the observations: the true energy of GRB $E_\gamma = 5\times
10^{50}ergs$ and the aspherical SNe of kinetic energy $E_{SN} =
2\times 10^{51}ergs$. For a set of GRBs the observed true energy
$E_\gamma $ and duration $T_{90} $ can be well fitted by taking
adequate values of the power-law index $n$ and magnetic field $B_H
$. However, there are still several issues related to this model.

(i) In this paper we take the evolving path of the \textbf{RP} in
the region \textbf{IA} to describe the association of GRB-SN as
shown in Figure 5. In this case the GRB will terminate before the
SN event stops. In fact the evolving path in region\textbf{ IB}
can be also used to describe the association of GRB-SN, and it
implies that the \textbf{RP} will arrive at the characteristic
curve of $\tilde {P}_{MC} = 0$ ahead of reaching the
characteristic curve of $\tilde {P}_{BZ} = 0$. However, we think
that the GRB will still terminate before the SN event stops,
because the disk will be destroyed during the explosion of the SN,
and the BZ process can not work without the magnetic field
supported by its surrounding disk.

(ii) In order to highlight the effects of the BZ and MC processes
in powering GRB-SN, we neglect the effects of the disk accretion
in this model. In fact, disk accretion plays a very important role
in BH evolution. The slowing down of the BH spin will be delayed
significantly since the accreting matter delivers remarkable
amount of energy and angular momentum to the BH. Strictly
speaking, disk accretion cannot be halted steadily by the MC
process. The MC effects on disk accretion is a very complicated,
and it should be treated dynamically.

(iii) According to Kruskal-Shafranov criteria (Kadomtsev 1966),
the screw instability will occur if the toroidal magnetic field
becomes so strong that the magnetic field lines turns around
itself about once. In our latest work (Wang et al. 2004), we
argued that the state of CEBZMC always accompanies the screw
instability. In this model the half-opening angle $\theta _{BZ} $
is related to the infinite radial parameter $\xi $ by equation
(\ref{eq4}). We expect to derive a greater $\theta _{BZ} $ by
considering the restriction of the screw instability to the MC
region. Thus the ratios $\eta _{GRB} $ and $\eta _{SN} $ will be
changed. We shall improve this model in future.

\section{Acknowledgments}
We thank the anonymous referees for numerous constructive
suggestions. This work is supported by the National Natural
Science Foundation of China under grants 10173004, 10373006 and
10121503.


\clearpage
\begin{deluxetable}{cccccc}
\tabletypesize{\scriptsize}
\tablecaption{Characteristic of BH
evolution in each region of the parameter space }
\tablewidth{0pt}

\tablehead{\colhead{Region} & \colhead{$da_* / dt$} &
\colhead{$P_{BZ} $} & \colhead{$P_{MC} $} & \colhead{\textbf{RP}
displacement}& \colhead{Possible events} }

\startdata

\textbf{IA} & $<0$ & $>0$ & $>0$ & Towards the left& GRB-SN \\

\textbf{IB} & $<0$ & $>0$ & $>0$ & Towards the left& GRB-SN \\

\textbf{II} & $<0$ & ------ & $>0$ & Towards the left& SN \\

\textbf{IIIA} & $<0$ & ------ & $<0$ & Towards the left& ------ \\

\textbf{IIIB} & $<0$ & $>0$ & $<0$ & Towards the left& GRB \\

\textbf{IVA} & $>0$ & ------ & $<0 $& Towards the right& ------ \\

\textbf{IVB} & $>0$ & $>0$ & $<0$ & Towards the right& GRB \\

\enddata
\end{deluxetable}

\begin{deluxetable}{cccccc}
\tabletypesize{\scriptsize}
\tablecaption{Four GRBs of the given
true energy $E_\gamma $ and $T_{90}$s fitted with the different
power-law index $n$ and the predicted $E_{SN}$ }
\tablewidth{0pt}

\tablehead{\colhead{GRB\tablenotemark{a}} & \colhead{$E_\gamma
(10^{51}ergs)$\tablenotemark{b}} & \colhead{$T_{90} $(s)} &
\colhead{$n$} & \colhead{$B_H (10^{15}G)$} &
\colhead{$E_{SN}$\tablenotemark{c}} }

\startdata

970508 & 0.234 & 15\tablenotemark{d} & 3.885 & 0.97 & 1.947$\times 10^{51}ergs(\textstyle{{\varepsilon _{SN} } \over {0.0035}})$ \\

990712 & 0.445 & 30\tablenotemark{e} & 3.975 & 0.81 & 1.989$\times 10^{51}ergs(\textstyle{{\varepsilon _{SN} } \over {0.0035}})$ \\

991208 & 0.455 & 39.84\tablenotemark{f} & 3.985 & 0.75 & 1.995$\times 10^{51}ergs(\textstyle{{\varepsilon _{SN} } \over {0.0035}})$ \\

991216 & 0.695 & 7.51\tablenotemark{f} & 4.058 & 1.80 & 2.032$\times 10^{51}ergs(\textstyle{{\varepsilon _{SN} } \over {0.0035}})$ \\

\enddata

\tablenotetext{a}{\textbf{The sample of GRB-SN takes from the
sample listed in Table 1 of Dar (2004)}}
\tablenotetext{b}{\textbf{The true energy $E_\gamma $ of GRB
refers to the sample listed in Table 1 of Frail et al (2002)}}
\tablenotetext{c}{\textbf{The predicted energy of SNe based on our
model}}
\tablenotetext{d}{\textbf{The duration of GRB970508 refers
to Costa et al (1997)}}
\tablenotetext{e}{\textbf{The duration of
GRB990712 refers to Heise et al. (1999)}}
\tablenotetext{f}{\textbf{The durations of GRB991208 and GRB991216
refer to the sample listed in Table 2 of Lee {\&} Kim (2002)}}

\end{deluxetable}

\begin{deluxetable}{ccc}
\tabletypesize{\scriptsize}
\tablecaption{Energy and duration
obtained in three models for GRBs with $a_\ast (0) = 0.9$\textbf{,
}$M_H (0) = 7M_ \odot $\textbf{, }$B_H = 10^{15}G$\textbf{ and the
initial disk mass }$M_D (0) = 3M_ \odot $ }
\tablewidth{0pt}

\tablehead{\colhead{Model} & \colhead{$E_\gamma (ergs)$} &
\colhead{$T_{90} $(s)} }

\startdata

BZO & $1.225\times 10^{53}$ & 348 \\

BZACC & $1.734\times 10^{53}$ & 397 \\

CEBZMC & $ < 1.050\times 10^{52}$ & $<115$ \\

\enddata
\end{deluxetable}

\clearpage
\begin{figure}
\epsscale{0.5} \plotone{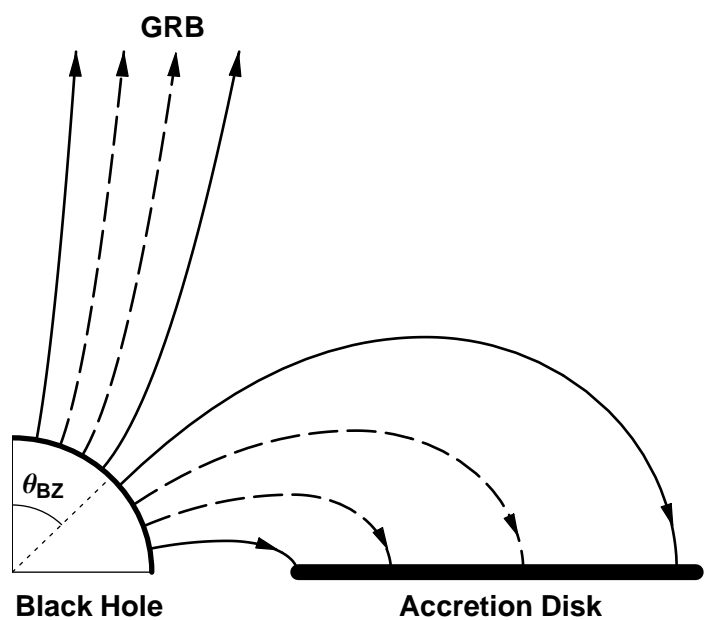} \figcaption{ Poloidal magnetic
field connecting a rotating BH with remote astrophysical load and
a surrounding disk}
\end{figure}

\clearpage

\begin{figure}
\epsscale{0.5}
\plotone{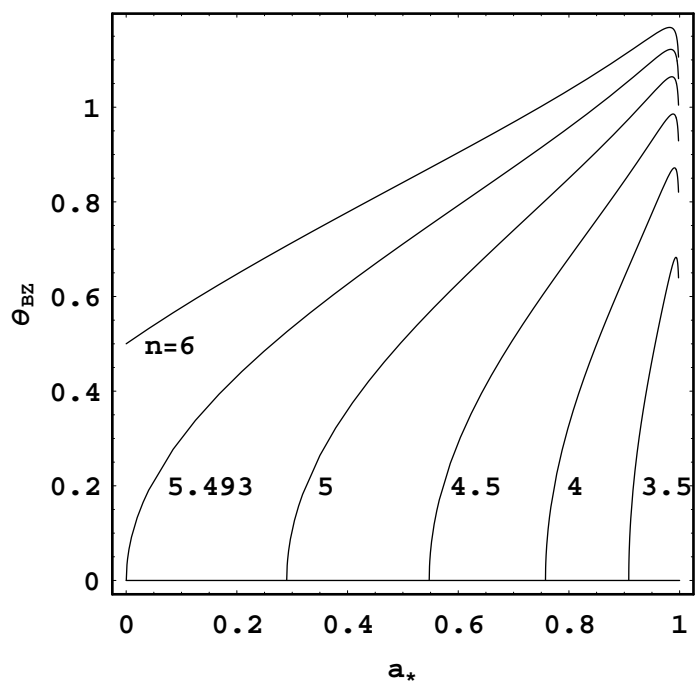}
\figcaption{Curves of $\theta_{BZ}
$ vs. $a_ * $  for $n=$3.5, 4, 4.5, 5, 5.493 and 6.}
\end{figure}

\clearpage

\begin{figure}
\epsscale{0.5}
\plotone{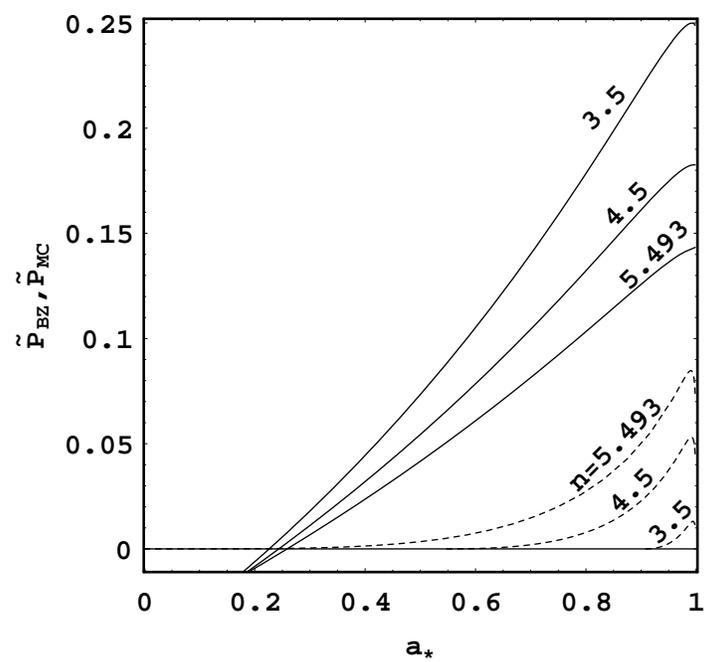}
\figcaption{Curves of $\tilde
{P}_{BZ} $(dotted lines) and $\tilde {P}_{MC} $(solid lines) vs.
$a_* $ with $n$=3.5, 4.5 and 5.493 for $0 < a_* < 1$.}
\end{figure}

\clearpage

\begin{figure}
\epsscale{0.5}
\plotone{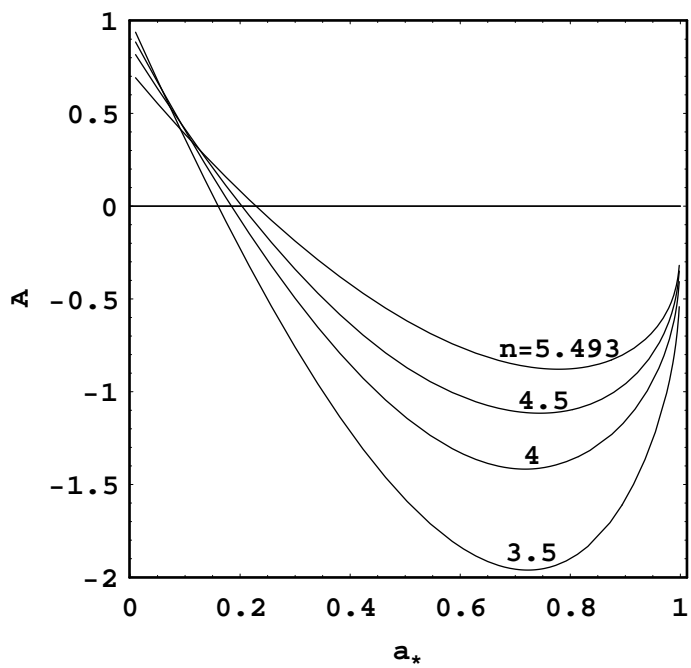}
\figcaption{Curves of $A(a_* ,n)$
vs. $a_\ast $ for $n$=3.5, 4, 4.5 and 5.493.}
\end{figure}

\clearpage

\begin{figure}
\epsscale{0.5}
\plotone{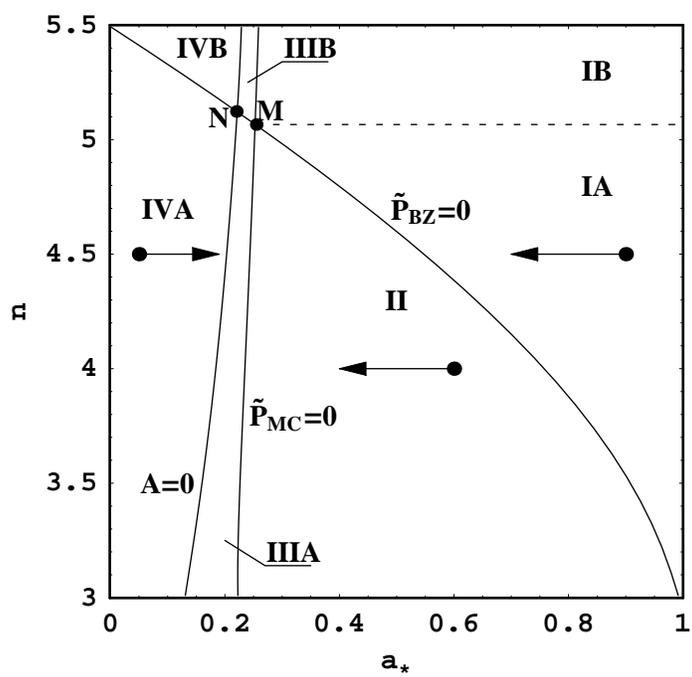}
\figcaption{Parameter space of BH
evolution in CEBZMC.}
\end{figure}

\clearpage

\begin{figure}
\epsscale{0.5}
\plotone{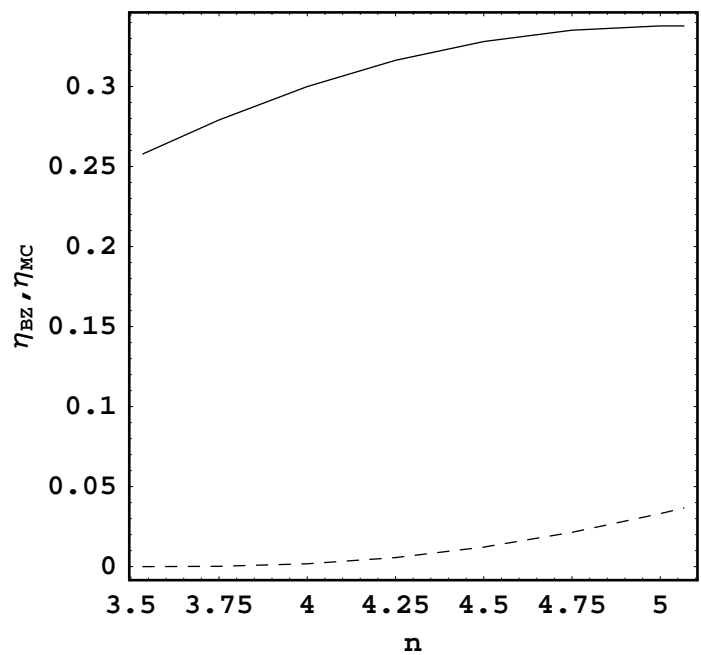}
\figcaption{Curves of $\eta _{BZ}
$(dashed line) and $\eta _{MC} $(solid line) vs. $n$ for $3.534 <
n < 5.067$.}
\end{figure}

\clearpage

\begin{figure}
\epsscale{0.5}
\plotone{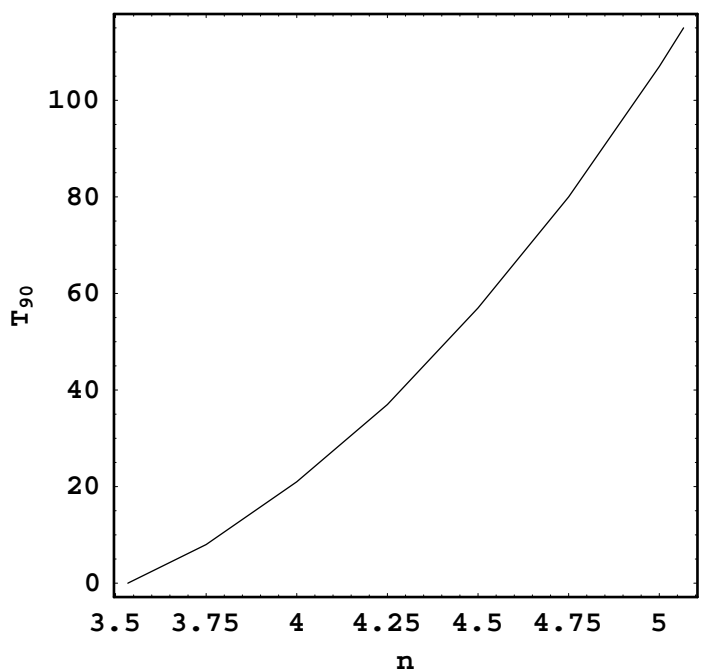}
\figcaption{Curve of $T_{90} $ vs.
$n$ for $3.534 < n < 5.067$.}
\end{figure}


\end{document}